\documentstyle[aps,pre,multicol,psfig]{revtex}
\begin{document}
\draft
           
\title{Reptation in the Rubinstein-Duke model: the influence of end-reptons dynamics}
\author{Enrico Carlon$^{1}$, Andrzej Drzewi\'nski$^{2}$ and 
J.M.J. van Leeuwen$^{3}$}
\address{
$^{1}$Theoretische Physik, Universit\"at des Saarlandes, 
D-66123 Saarbr\"ucken, Germany \\
$^{2}$Institute of Low Temperature and Structure Research, Polish Academy 
of Sciences, P.O. Box 1410, 50-950 Wroc\l aw 2, Poland\\
$^{3}$Instituut-Lorentz, University of Leiden, P.O. Box 9506, 
2300 RA Leiden, The Netherlands 
}
\date{\today}
\maketitle

\begin{abstract}
We investigate the Rubinstein-Duke model for polymer reptation by means of
density-matrix renormalization group techniques both in absence and presence 
of a driving field. In the former case the renewal time $\tau$ and the
diffusion coefficient $D$ are calculated for chains up to $N=150$ reptons and 
their scaling behavior in $N$ is analyzed. 
% Both quantities scale as powers of 
% $N$: $\tau \sim N^z$ and $D \sim 1/N^x$ with $2.7 <z< 3.3$ and $1.8 <x< 2.1$ 
% for the parameters considered, while asymptotically they tend to the 
% theoretical values $z=3$ and $x=2$. These effective exponents are sensitive 
% to the dynamics of the end reptons and we suggest how to influence this 
% dynamics and corresponding experiments to bring out such behavior.
Both quantities scale as powers of $N$: $\tau \sim N^z$ and $D \sim 1/N^x$ 
with the asymptotic exponents $z=3$ and $x=2$, in agreement with the reptation
theory. For an intermediate range of lengths, however, the data are well-fitted
by some effective exponents whose values are quite sensitive to the dynamics 
of the end reptons. We find $2.7 <z< 3.3$ and $1.8 <x< 2.1$ for the range of
parameters considered and we suggest how to influence the end reptons dynamics
in order to bring out such a behavior. 
At finite and not too small driving field, we observe the onset of the
so-called band inversion phenomenon according to which long polymers migrate
faster than shorter ones as opposed to the small field dynamics.
For chains in the range of 20 reptons we present detailed shapes of the 
reptating chain as function of the driving field and the end repton dynamics.
\end{abstract}

\begin{multicols}{2} \narrowtext
\section{Introduction}

The idea of reptation was introduced about thirty years ago by de Gennes
\cite{deGe71} in order to explain some dynamical properties of polymer 
melts of high molecular weight. 
The dynamics of such systems is strongly influenced by entanglement
effects between the long polymer chains. The basic idea of reptation is
that each polymer is constrained to move within a topological tube due to 
the presence of the confining surrounding polymers \cite{deGe79,Doi89}.
Within this tube the polymer performs a snake-like motion and advances in 
the melt through the diffusion of stored length along its own contour. 
One focuses thus on the motion of a single test chain, while the rest of 
the environment is considered frozen, i.e. as formed by a network of fixed 
obstacles.
The reptation theory predicts that the viscosity $\mu$ and longest relaxation 
time $\tau$ (known as {\it renewal} time) scale as $\mu \sim \tau \sim N^3$, 
where $N$ is the length of the chains, while the diffusion constant scales as 
$D \sim 1/N^2$.
These results are not far from the experimental findings. Measurements of 
viscosity of concentrated polymer solutions and melts of different chemical 
composition and nature are all consistent with a scaling $\mu \sim N^{3.4}$
\cite{Ferr80}, while for the diffusion constant both $D \sim 1/N^2$ 
\cite{Klei78} and $D \sim 1/N^{2.4}$ \cite{Kim86,Nemo89,Lodg99,Thao00}
have been reported. This discrepancy triggered a 
substantial effort in order to reconcile theory and experiments.

Another physical situation where reptation occurs is in {\it gel 
electrophoresis}, where charged polymers diffuse through the pores of a gel
under the influence of a driving electric field \cite{Viov00}. The gel 
particles form a frozen network of obstacles in which the polymer moves 
through the diffusion of stored length. Electrophoresis has important 
practical applications, for instance in DNA sequencing, since it is a 
technique which allows to separate polymers according to their length.

The previous examples demonstrate how reptation is an important
mechanism for polymer dynamics in different physical situations.
A prominent role in understanding the subtle details of the dynamics was
played by models defined on the lattice, which besides describing correctly
the process at the microscopic level, offer important computational advantages.
The aim of this paper is to investigate in detail one of such lattice models,
which was originally introduced by Rubinstein \cite{Rubi87}
and later extended by Duke \cite{Duke89} in 
order to include the effect of an external driving field.
The Rubinstein - Duke (RD) model has been studied in the past using 
several techniques; there are a limited numbers of exact results available
\cite{Wido91,vanL91,Prah96}
while a rich literature on simulation results on the model exists.
The latter have been mostly obtained by Monte Carlo (MC) simulations 
\cite{Rubi87,Duke89,Bark94,Bark97}. MC simulations are tedious for long chains 
since the renewal time scales as $N^3$, thus it is difficult to obtain a small 
statistical error for large $N$.

In this paper we study the RD model by means of Density Matrix Renormalization 
Group (DMRG) \cite{Whit92}, a technique which has been quite successful in 
recent years, in particular in application to condensed matter problems as 
quantum spin chains and low-dimensional strongly correlated systems
\cite{DMRGbook}. 
Using the formal similarity between the Master equation for reptation and the
Schr\"odinger equation, DMRG also allows to calculate accurately stationary 
state properties of long reptating chains.

Some of the results reported here, in particular concerning the scaling of
the renewal time $\tau$, have been presented before \cite{Carl01a}. Here 
we will give full account of the details of the calculations and present a 
series of new results concerning reptation in the presence of an electric 
field.
One of the main conclusions of our investigation is that the exponents 
describing the scaling of $\tau$ and $D$ in the intermediate length region
appear to be rather sensitive on the structure of the end repton. 
We find that, by influencing the end repton dynamics, one has
a regime where the effective exponents are considerably {\it lower} than the
standardly found experimental values. 
Experiments are suggested, involving polymer architectures not investigated 
so far, and which ought to confirm our predictions.

This paper is organized as follows: in Sec. \ref{sec:repton} we introduce
the RD model, while in Sec. \ref{sec:dmrg} we briefly outline the
basic ideas of DMRG. Sec. \ref{sec:zero_field} is dedicated to the properties
of reptation in absence of an external field, in particular to the scaling
properties of $\tau$ and $D$.
Sec. \ref{sec:field} collects a series of results of reptation in a field, 
while Sec. \ref{sec:discussion} concludes our paper.

\section{The Rubinstein - Duke model}
\label{sec:repton}

\begin{figure}[b]
\centerline{\psfig{file=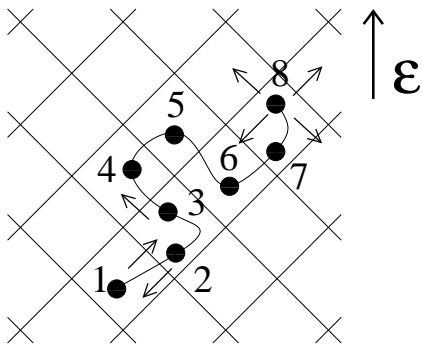,height=4.5cm}}
\vskip 0.2truecm
\caption{Example of a configuration in the RD model in $d=2$ dimensions.
In terms of the relative coordinates projected along the field direction
the configuration reads $y=\{ 1,0,1,1,-1,1,0\}$. A non-zero field 
($\varepsilon$) biases the motion of the reptons, which occurs with 
rates $B=\exp(\varepsilon/2)$ $(B^{-1})$ for moves in the direction 
(opposite to) the applied field. 
The arrows indicate the possible moves. Notice that an end repton can
stretch to $d$ lattice positions forward and backward in the field
(as repton $8$ in the example).
}
\label{FIG01}
\end{figure}

In the RD model the polymer is divided into $N$ units, called reptons, which 
are placed at the sites of a $d$-dimensional hypercubic lattice (see Fig.
\ref{FIG01}). The number of reptons that each site can accommodate is unlimited 
and self-avoidance effects are neglected.
Each configuration is projected onto an axis along the (body)
diagonal of the unit cell and it is identified by the relative coordinates 
$y_i \equiv z_{i+1} - z_{i}$ of neighboring reptons along the chain ($z_i$ 
indicates the projected coordinate of the $i$-th repton). The relative 
coordinates can take three values $y_i =-1,0,1$ and there are thus in total 
$3^{N-1}$ different configurations for a chain with $N$ reptons.

When two or more reptons accumulate at the same lattice site they form a part 
of stored length, which can then diffuse along the chain. In terms of relative
coordinates a segment of stored length corresponds to $y_i=0$, therefore 
allowed moves are interchanges of $0$'s and $1$'s, i.e. $0,\pm 1 
\leftrightarrow \pm 1,0$. 
On the contrary, the end reptons of the chain can stretch ($0 \to \pm 1$) or retract to
the site occupied by the neighboring repton ($\pm 1 \to 0$).
The dynamics of the chain is fully specified once the rates for the moves are
given. A field $\varepsilon$ along the projection axes is introduced so that
moves forward and backward in the field occur with rates $B = \exp 
(\varepsilon/2)$ and $B^{-1} = \exp (-\varepsilon/2)$. In the following we will
be interested in both cases $\varepsilon = 0$ and $\varepsilon > 0$.
Notice that when an end repton moves towards an empty lattice site (tube 
renewal process) it has in total $d$ different possibilities of doing so
forward and backward in the field (see Fig. \ref{FIG01}), therefore the 
associated rates
are $d B$ and $d B^{-1}$ respectively. On the contrary the end repton can 
retract by moving to the (unique) site occupied by its neighbor with rate
$B$ or $B^{-1}$.
Summarizing the possible moves are: (a) stored length diffusion for inner 
segments, i.e. ($0,\pm 1 \leftrightarrow \pm 1, 0$) with rates $B$ and 
$B^{-1}$, (b) contractions for external reptons ($\pm 1 \to 0$) with rates
$B$ and $B^{-1}$ and (c) stretches for external reptons ($0 \to \pm 1$), with
rates $d B$ and $d B^{-1}$.
Notice that the parameter $d$ enters in the model only at the stretching rates
(c). Rather than linking $d$ to the lattice coordination number we interpret 
it as the ratio between stretching rates and rates associated to moves of 
inner reptons. This allows to choose any positive values for $d$.

\begin{figure}[b]
\centerline{\psfig{file=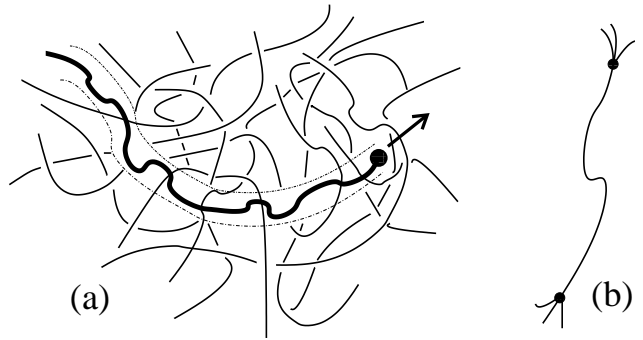,height=4.5cm}}
\vskip 0.2truecm
\caption{
(a) In thick line we indicate a test chain moving in an environment of other 
chains. By modifying its ends, for instance, by attaching molecules of
large size one can lower the parameter $d$, as in this way one expects that
stretchings of the chain out of the tube will occur to lower rate.
(b) The lowering of $d$ would also occur for reptating polymers with short
branching ends.
}
\label{FIG02}
\end{figure}

As we will see, a variation of $d$ has some important effects on the 
corrections to the asymptotic behavior. This regime may
turn out to be relevant for the experiments. Moreover it is conceivable that
the parameter $d$ could be somewhat modified in an experiment. Figure 
\ref{FIG02} schematically illustrates a possibility of doing so, namely
by attaching to the chain ends some large molecules so that for each chain
tube renewal moves would be impeded. The same effects would be possible if
the linear chain is modified such as to have branches close to the endpoints
(Fig. \ref{FIG02}(b)), or an end part stiffer than the rest of the chain
(as it could be realized in block copolymers).
In both cases we expect that stretches out of the confining tube would be 
suppressed while chain retractions would not be much impeded.

Once the rates for elementary processes are given, the stationary properties 
of the system can be found from the solution of the Master equation
\begin{eqnarray}
\frac{d P(y,t)}{dt} 
&=&  - \sum_{y'} H_{yy'} P(y',t)
\label{mastereq}
\end{eqnarray}
in the limit $t \to \infty$. Here $P(y,t)$ indicates the probability of 
finding the polymer in a configuration $y$ at time $t$ and the matrix
$H$ contains the transition rates per unit of time between the different 
configurations of the chains, as given in the rules discussed above.

\section{Density matrix renormalization}
\label{sec:dmrg}

DMRG was introduced in 1992 by S. White \cite{Whit92} as an efficient 
algorithm to deal with a quantum hamiltonian for one dimensional systems. 
It is an iterative basis--truncation method, which allows to approximate 
eigenvalues and eigenstates, using an optimal basis of size $m$. It is not 
restricted to quantum system; it has also been successfully applied to a 
series of problems, ranging from two dimensional classical systems 
\cite{class_2d} to stochastic processes \cite{Carl99}. The basic common 
feature of these problems is the formal analogy with the Schr\"odinger 
equation for a one dimensional many--body system. DMRG can also be applied 
to the Master Equation (\ref{mastereq}) for the reptating chain, albeit with 
some limitations due to the non--hermitian character of the matrix $H$. In 
this specific case we start from small chains, for which $H$ can be 
diagonalized completely, and construct effective matrices representing 
$H$ for longer chains through truncation of the configurational space followed 
by an enlargement of the chain. This is done through the construction of a 
reduced density matrix whose eigenstates provide the optimal basis set, as
can be shown rigorously (see Refs. \cite{Whit92,DMRGbook} for details).
The size $m$ of the basis remains fixed in this process. By enlarging $m$ one 
checks the convergence of the procedure to  the desired accuracy. Hence $m$ 
is the main control parameter of the method.
In the present case we found that $m=27$ is sufficient for small driving 
fields and we kept up to $m=81$ state for stronger fields.

$H$ is only hermitian for zero driving field. So for a finite driving field
one needs to apply the non--hermitian variant of the standard DMRG algorithm
\cite{Carl99}.
For a non--hermitian matrix $H$ one has to distinguish between the right 
and left eigenvector belonging to the same eigenvalue. Since $H$ is a 
stochastic matrix the lowest eigenvalue equals 0 and the corresponding left 
eigenvector is trivial (see next section). The right eigenvector gives the 
stationary probability distribution. The next to lowest eigenvalue, the gap 
$1/\tau$, yields the slowest relaxation time $\tau$ of the decay towards the 
stationary state. Generally the DMRG method works best when the eigenvalues 
are well separated. For long chains and stronger driving fields the spectrum 
of $H$ gets an accumulation of eigenvalues near the zero eigenvalue of the 
stationary state. This hampers the convergence of the method seriously and 
enlarging the basis $m$ becomes of little help, while standardly this 
improves the accuracy substantially.
In order to construct the reduced density matrix from the lowest eigenstates
one needs to diagonalize the effective matrices $H$ at each DMRG step, we used
the so-called Arnoldi method which is known to be particularly 
stable for non-hermitian problems \cite{Arnoldi}.

\section{Zero field properties}
\label{sec:zero_field}

In this section we present a series of DMRG results on the scaling behavior
as function of the polymer length $N$ of the so-called tube renewal time
$\tau(N)$, i.e. of the characteristic time for reptation, and of the 
diffusion coefficient $D(N)$.

\subsection{Tube renewal time}
\label{subsec:renewaltime}

The renewal time, $\tau$, is the typical time of the reptation process, i.e. 
the time necessary to loose memory of any initial configuration through 
reptation dynamics.
$\tau$ is given by the inverse of the smallest gap of the matrix $H$, which
can be seen as follows: Starting from an initial configuration
%with probability distribution $| P_0 \rangle$, 
one has, asymptotically for 
sufficiently long times ($t \to \infty$):
\begin{equation}
| P(t) \rangle = | \phi_0 \rangle + e^{-t/\tau} | \phi_1 \rangle + \ldots
\label{asymptotic}
\end{equation}
where $|\phi_0 \rangle$ and $|\phi_1 \rangle$ are eigenstates of the matrix $H$
defined in Eq. (\ref{mastereq}) with eigenvalues $E_0 = 0$ and $E_1 = 1/\tau$,
respectively. The state $|\phi_0 \rangle$ is the stationary state of the 
process, while the gap of $H$ corresponds to the inverse relaxation time.
Notice that, as the matrix $H$ is symmetric at zero field,
$E_1$ is always a real number, while for non-Hermitian matrices the eigenvalues
may get an imaginary part, which causes a relaxation to equilibrium through 
damped oscillations.

An important point of which we took advantage in the calculation is the fact 
that the ground state $| \phi_0 \rangle$ of the matrix $H$ is known exactly 
in absence of any driving fields. Since $H$ is stochastic and symmetric (i.e. 
$\sum_{y'} H_{y y'} = \sum_{y'} H_{y' y} = 0$), it follows immediately 
that the stationary state is of the form $\phi_0 (y) = c$, with $c$ a constant.
Instead of applying the DMRG technique directly to the matrix $H$ we applied it
to the matrix $H'$ defined by:
\begin{equation}
 H' = H + \Delta | \phi_0\rangle \langle \phi_0|.
\end{equation}
Now if we choose $\Delta > E_1$ the lowest eigenvalue of $H'$ is equal to
$E_1$, therefore the problem of calculating the gap of $H$ is reduced to
the calculation of the ground state of a new (non-stochastic) matrix $H'$. 
This approach is considerably more advantageous in term of CPU time and memory 
required for the program (for more details see Ref. \cite{Carl01b}).

\begin{figure}[b]
\centerline{\psfig{file=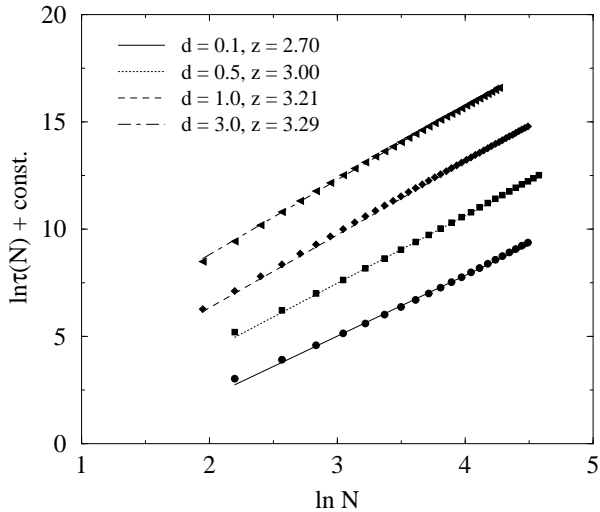,height=7.cm}}
\vskip 0.2truecm
\caption{Plot of $\ln \tau (N)$ vs. $\ln N$ for various $d$. We report the 
values of the slope of the data obtained from a linear interpolation.}
\label{FIG03}
\end{figure}

Figure \ref{FIG03} shows a plot of $\ln \tau(N)$ vs. $\ln N$ for the renewal
time as calculated from DMRG methods for various values of the stretching 
rate $d$ and for lengths up to $N=150$. The data have been shifted along the 
abscissae axis by an arbitrary constant. 
We fitted the data by a linear interpolation, as it is mostly done in Monte
Carlo simulations and in experiments. The resulting slopes provide estimates
of the exponent $z$, which we find to be sensitive to the stretching rate $d$.
For $d$ sufficiently large ($d > 2$) we find $z \approx 3.3$, in 
agreement with experiments \cite{Ferr80} and previous Monte Carlo simulation 
results \cite{Rubi87,Deut89}.
The value of $z$ decreases when $d$ is decreased. For $d$ sufficiently small
($d \approx 0.1$) we find $z \approx 2.7 < 3$, which is a regime that has not
yet been observed in experiments. 

To shed some more light into these results we considered the {\it effective 
exponent}, i.e. the following quantity: 
\begin{equation}
z_N = \frac{\ln \tau (N+1) - \ln \tau (N-1)}{\ln (N+1) - \ln (N-1)} \,\, ,
\label{zN}
\end{equation}
which is the discrete derivative of the data in the log-log scale plot.
Such quantity probes the local slope (at a given length $N$) of the data of
Fig. \ref{FIG03} and provides a better estimate of the finite $N$ corrections 
to the asymptotic behavior.
The calculation of $z_N$ requires very accurate data and cannot be easily
performed by Monte Carlo simulation results which are typically affected by numerical
uncertainties, as these are amplified when taking numerical derivatives.
We stress that already in the log-log plot of Fig. \ref{FIG03} the deviation 
of the data from linearity is noticeable, therefore the values of $z$ given 
above are just average values and not to be expected as asymptotic ones.

\begin{figure}[b]
\centerline{\psfig{file=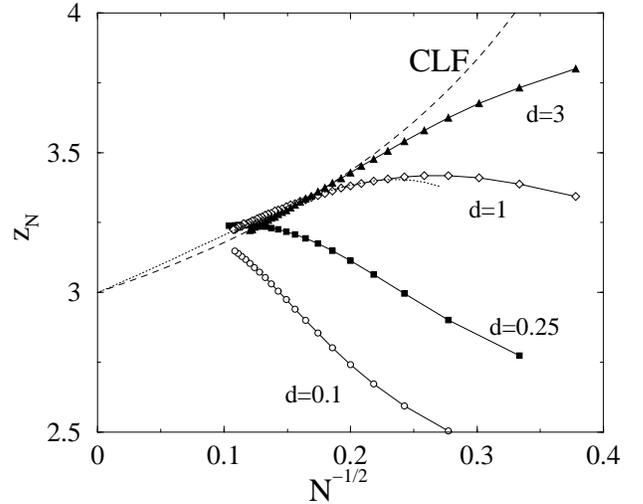,height=7.cm}}
\vskip 0.2truecm
\caption{Solid lines: Effective exponent $z_N$ for various $d$. Dashed line: 
Corrections due to CLF as predicted by Doi (Eqs. \ref{doi} and \ref{effdoi})
for the $d=3$ curve. Dot-dashed line: Inclusion of higher order corrections
as given by Eq. \ref{effMM} for the curve $d=1$.}
\label{FIG04}
\end{figure}

Figure \ref{FIG04} shows a plot of $z_N$ for the data given in 
Fig. \ref{FIG03}, plotted as function of $1/\sqrt N$. In the thermodynamic 
limit $N \to \infty$, $z_N$ is seen to converge towards $z=3$, in agreement
with de Gennes' theory \cite{deGe71}. Corrections to this asymptotic limit 
yield $z_N > 3$
when $N$ is sufficiently large, with deviations towards $z_N < 3$ for small 
$d$ and not too large $N$.

There exist some theoretical predictions for the form of the corrections
to the asymptotic behavior of the renewal time \cite{Doi81}. These are based 
on contour length fluctuations (CLF), i.e. on the idea that the length of the 
tube fluctuates in time and that this would help accelerating the renewal 
process.
Such fluctuations were not taken into account in the original work of 
de Gennes, who assumed the tube to have a fixed length.
According to CLF theory, thus, $\tau$ scales as \cite{Doi81}
\begin{equation}
\tau (N) \sim N^3 \left( 1 - \sqrt{\frac{N_0}{N}} \,\, \right) ^2,
\label{doi}
\end{equation}
with $N_0$ a typical length.
Substituting this equation into Eq. (\ref{zN}) one obtains: 
\begin{equation}
z_N^{\rm (CLF)} = 3 + \frac{\sqrt{N_0/N}}{1-\sqrt{N_0/N}}\,\, .
\label{effdoi}
\end{equation}
which is a monotonically increasing function of $1/\sqrt N$. For all lengths 
in the physical range $N > N_0$ one has $z_N^{\rm (CLF)} > 3$. 
Eq. (\ref{effdoi}) is plotted as a dashed line in Fig. \ref{FIG04}, where 
we have chosen $N_0 = 2.3$ in order to obtain the best fit of the DMRG 
data with $d=3$. This value is consistent with that predicted by the CLF
theory \cite{Doi89}. Our results further suggest that the value of
$N_0$ increases when $d$ is decreased.

While for sufficiently long chains the data apparently merge to the CLF 
theory given by Eq.(\ref{effdoi}) higher
order corrections appear to be of opposite sign. When $d$ is lowered the latter
become particularly strong so that the effective exponent for a certain range 
of lengths is even smaller than $3$. To our knowledge such an effect has not 
yet been observed in experiments. Presumably standard polymer mixtures will 
correspond to $d \gtrsim 1$, which is a regime where an exponent $z \approx
3.4$ is observed \cite{Ferr80}. It is conceivable that mixtures with modified 
architecture, as those illustrated in Fig. \ref{FIG02}(b), i.e. long polymers 
with short branching ends, would correspond to curves at much lower $d$. 
Experimental results for such systems would be of high interest in order to 
check the predictions of the RD model in the low $d$ regime.

In order to gain some more insight in the precise form of $\tau (N)$ we 
considered higher order terms which lead to an expression of the type:
\begin{equation}
\tau(N) \sim N^3 \left[ \left( 1 - \sqrt{\frac{N_0}{N}} \right)^2 + 
A \frac{N_0}{N} \right]
\end{equation}
Recently, Milner and McLeish \cite{Miln98} formulated a more complete theory 
beyond that of CLF by Doi, using ideas from the theory of stress relaxation 
for star polymers \cite{Miln97}, which to lowest orders in $1/\sqrt{N}$ yields 
an expression of the type given above. The effective exponent now reads:
\begin{equation}
z_N = 3 + \sqrt{\frac{N_0}{N}}
\frac{1 - (1+A)\sqrt{{N_0}/{N}} }
{\left( 1 - \sqrt{{N_0}/{N}} \right)^2 + A {N_0}/{N}}
\label{effMM}
\end{equation}
Notice that the CLF expression (Eq. \ref{effdoi}) diverges for $N \to N_0$, 
while the previous formula this divergence does not occur. For $N$ not too 
large the numerator in Eq. (\ref{effMM}) may change sign, reproducing 
features found in the DMRG calculations, i.e. an effective exponent $z < 3$.
The dot-dashed line of Fig. \ref{FIG04} represents a fit for the $d=1$ case 
using Eq.(\ref{effMM}); we find that the choice $N_0 = 3.6$ and $A = 0.44$ 
fits very well the numerical data for $N > 25$, while deviations for shorter 
chains are clearly visible.
We stress that Eq. (\ref{effMM}) fits quite well the renewal time data in the 
cage model \cite{Faso01}, which is another lattice model of reptation dynamics 
\cite{Evan81}.
Finally, very recently the effect of constraints release (CR), introduced in 
the RD model in a self-consistent manner, has been considered \cite{Paes02}. 
Apart from small quantitative shifts in the effective exponents, the 
is unchanged.

\subsection{Diffusion coefficient}
\label{subsec:diffusion}

We consider now a similar scaling analysis of the diffusion coefficient $D$.
To compute the diffusion constant as function of the chain length we
used the Nernst--Einstein relation \cite{vanL91}:
\begin{equation}
D = \lim_{\varepsilon \to 0} \frac{v}{N \varepsilon}.
\label{NE}
\end{equation}
where $v$ is the drift velocity of the polymer with $N$ reptons and subject to
an external field $\varepsilon$.
In order to estimate $D(N)$ we considered small fields (down to $\varepsilon
\sim 10^{-5}$) and calculated the velocity $v_i$ of the $i$-th repton in the
stationary state using the formulas given in Section \ref{sec:fieldb}.
As a check for the accuracy of the procedure we verified that the drift 
velocity $v_i$ is independent on $i$, i.e. the position of the repton along 
the chain in which it is computed.

%==============================================================================
\vbox{
\begin{table}[tb]
\caption{Comparison between diffusion coefficients from exact diagonalization
methods, as given in Ref. \protect\cite{Bark97}, and as obtained from 
the calculation of $r(\varepsilon)=v/N \varepsilon$ from the DMRG algorithm
with $m=18$ for decreasing $\varepsilon$.}
\vskip 0.25truecm
\label{TABLE01}
\begin{tabular}{ccccc}
$N$ & $r(\varepsilon=10^{-2})$  & $r(\varepsilon=10^{-3})$  &
$r(\varepsilon=10^{-4})$  & $DN^2$ (Ref. \protect\cite{Bark97})
\\
\hline
 5 & 0.864873 &  0.864907 & 0.864908 & 0.864908 \\
 7 & 0.769024 &  0.769057 & 0.769057 & 0.769057 \\
 9 & 0.703975 &  0.703951 & 0.703951 & 0.703951 \\
11 & 0.657701 &  0.657550 & 0.657549 & 0.657549 \\
13 & 0.623372 &  0.623010 & 0.623006 & 0.623006 \\
15 & 0.596975 &  0.596304 & 0.596297 & 0.596297 \\
17 & 0.576097 &  0.575007 & 0.574996 & 0.574996 \\
19 & 0.559224 &  0.557595 & 0.557580 & 0.557579 \\
\end{tabular}
\end{table}
}
%=========================================================================== 

\begin{figure}[b]
\centerline{\psfig{file=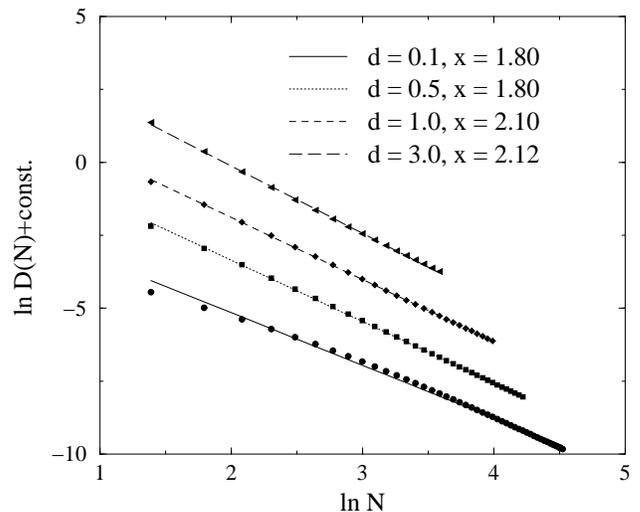,height=7.cm}}
\vskip 0.2truecm
\caption{Log-log plot of the diffusion constant as function of the chain
length for various values of the stretching rate $d$.}
\label{FIG05}
\end{figure}

In practice we estimated the ratio $r = v N/\varepsilon$ for decreasing
$\varepsilon$ until convergence was reached (typically we considered 
$\varepsilon \approx 10^{-4} - 10^{-5}$). Table \ref{TABLE01} shows the 
numerical values of $r(\varepsilon)$ for various $\varepsilon$ obtained from
DMRG with $m=18$ states kept, for short chains, so that they can be
compared with the exact diagonalization data shown in the last column.
The results for $\varepsilon = 10^{-4}$ are in excellent agreement with exact
diagonalization data for $D N^2$ (from Ref. \cite{Bark97}) reported in
the last column, which indicates that the procedure used to extrapolate the
diffusion coefficient from the Nernst-Einstein relation (Eq. \ref{NE}) is
correct and reliable.

According to reptation theory the diffusion coefficient for large $N$ should
scale as $D(N) \sim 1/N^2$ \cite{deGe71}. This results has been also derived 
rigorously for the RD model where the coefficient of the leading term is also 
known \cite{vanL92,Prah96}:
\begin{equation}
D(N) = \frac{1}{(2 d + 1) N^2}
\end{equation}

Early experimental results on diffusion coefficients for polymer melts were 
consistent with a power $-2$ \cite{Klei78}, while more recent results suggest
that such exponent would be significantly higher \cite{Lodg99}.
In fact the original measurements consistent with a scaling $1/N^2$ raised
quite some problems in the past, which was pointing to an inconsistence between
the scaling of $D$ and $\tau$ due to the following argument. For a 
reptating polymer, $D$ and $\tau$ the typical radius of gyration
should scale as:
\begin{equation}
R_g \sim \sqrt{D \tau}
\end{equation}
Now as the polymer obeys gaussian statistics $R_g \sim \sqrt{N}$, which 
implies that, if $\tau \sim N^{3.4}$ then necessarily $D \sim N^{-2.4}$,
for the same range of lengths. This argument is however not quite correct
in this form since $R_g \sim \sqrt{N}$ only asymptotically, thus
finite size corrections may affect $D$ and $\tau$ differently, as indeed
happens in the RD model.

\begin{figure}[b]
\centerline{\psfig{file=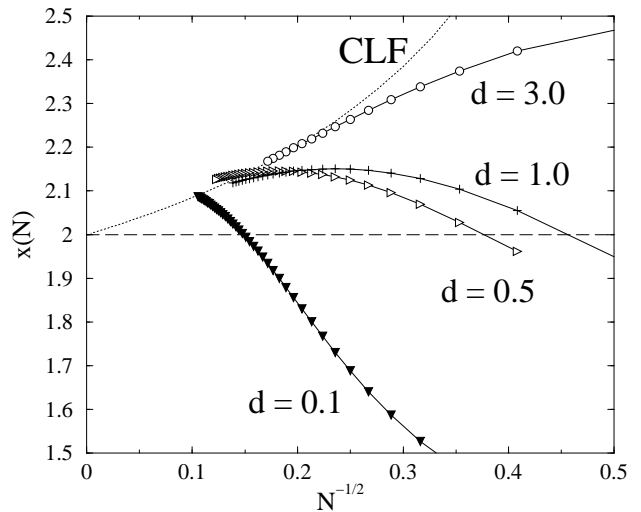,height=7.cm}}
\vskip 0.2truecm
\caption{Effective exponent $x_N$ as defined by Eq. (\ref{xN}) for various
$d$. The dotted line is the theoretical prediction from Eq. (\ref{CLF_D}).}
\label{FIG06}
\end{figure}

Figure \ref{FIG05} shows a plot of $\ln D$ as a function of $\ln N$ for various
values of the stretching rate $d$. The best fitting parameters $x$ for the
scaling $D \sim 1/N^x$ are given. As for the diffusion constant the
exponent passes the presumed asymptotic value of $2$ for $d=1$ and $d=3$,
while $x <2$ for $d=0.5$ and $d=0.1$.
Other numerical investigations of the
RD and related model yielded $x \approx 2.0$ \cite{Rubi87}, $x \approx 2.5$
\cite{Deut89}, and $x \approx 2.0$ \cite{Bark98}.
Again, it is best to analyze the effective exponent
\begin{equation}
x_N = -\frac{\ln D (N+1) - \ln D (N-1)}{\ln (N+1) - \ln (N-1)} \,\, ,
\label{xN}
\end{equation}
which is shown in Fig. \ref{FIG06}.
$x_N$ shows a similar behavior as the renewal exponent. However, comparing 
the same values of $N$ and $d$ in Figs. \ref{FIG04} and \ref{FIG06} one
notices that finite $N$ corrections are weaker for the diffusion coefficient.
For instance, for $d=3$ and $N^{-1/2} = 0.3$ one has $z_N \approx 3.8$, while
$x_N \approx 2.3$.
As mentioned above, recent experimental investigation of diffusion coefficient
for polymer melts and solutions yielded some different results concerning
its scaling behavior. Early measurements of polymer melts yielded 
$D \sim N^{-2}$ \cite{Klei78}, while in concentrated solutions typically
$D \sim N^{-2.4}$ \cite{Kim86,Nemo89}. 
Very recently, however, on hydrogenated polybutadiene concentrated 
solutions and melts it was found $D \sim N^{-2.4}$ for both cases 
\cite{Thao00}. A reanalysis of previous experimental results on several
different polymers lead to the conclusion $D \sim N^{-2.3}$ \cite{Lodg99}.
Thus the issue experimentally has not yet fully settled.
A theoretical analysis of the contour length fluctuations on the diffusion
coefficient was recently performed \cite{Fris00} leading to the expression:
\begin{equation}
D_{\rm CLF} (N) \sim N^{-2} \left( 1- \sqrt{\frac{N_0}{N}} \right)^{-1}
\label{CLF_D}
\end{equation}
In Fig. \ref{FIG06} we plotted the corresponding effective exponent, fitted
to the $d=3$ results. Again, the numerical results seem to approach the
CLF formula only for very long chains, where the effective exponent is
already quite close to the asymptotic value.
Our results indicate some variation of the effective exponent as function of 
the parameter $d$. It would be therefore interesting to investigate the 
scaling of $D$ for the polymer with short branching ends, as those 
illustrated in Fig. \ref{FIG02}(b) in order to test the validity of our 
predictions at smaller $d$. 

Finally we mention that the further order correction terms for the scaling
of $D$ have raised recently some debate about the nature of the expansion
\cite{Bark94,Prah96}. 
The DMRG results, which are accurate enough to investigate the higher
orders, clearly show that these results can be very well represented by a
series in $1/\sqrt{N}$ \cite{Carl01a}.

\section{Reptation in the presence of an electric field} 
\label{sec:field}

For finite values of $\epsilon$ severe limitations appear preventing 
us to extend the calculations to the long chains which we were able analyze
in the limit $\epsilon \rightarrow 0$. These limitations are intrinsic to the 
problem and the Arnoldi routine for finding the lowest eigenvalue of the
matrix H.
For finite $\epsilon$ and longer chains a massive accumulation of small 
eigenvalues near the ground state eigenvalue starts  to emerge. The problem  
is also present in the straight application of the routine to small chains,
for which an exact diagonalization of the matrix $H$ can be performed. 
For $N > 11$ and $\epsilon \geq 1$ the method for finding the 
lowest eigenvalue is no longer convergent. The DMRG method, described above, 
yields convergent results up to the chain $N=15$. 
For smaller fields, $\varepsilon < 1$, which is however the most interesting 
region from the physical viewpoint, the calculation can be extended to 
somewhat longer chains i.e. $N \approx 30-40$.
Neither extension of the truncated basis set, nor the 
inclusion of more target states is a remedy for the lack of convergence. 
Despite the intrinsic difficulty of treating long chains, substantial
information can be obtained by the analysis of the drift velocity for
intermediate chain lengths as function of $\varepsilon$.

\subsection{Drift velocity vs. field}
\label{sec:fielda}

\begin{figure}
\centerline{
{\psfig{file=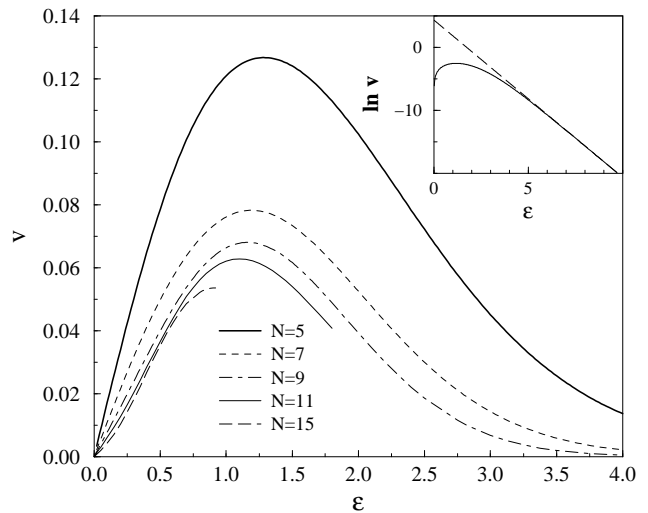,height=7.0cm}}
}
\vskip 0.4truecm
\caption{The velocities as a function of an electric field for chains of 
different lengths for $d=1$. Inset: $\ln v$ vs. $\varepsilon$ for $N=7$; the
asymptotic behavior is consistent with the prediction from Eq.(\ref{kolo}),
shown as a dashed line.
}
\label{FIG07}
\end{figure}

In Fig. \ref{FIG07} we have plotted the drift velocity as function of $\epsilon$ 
for chains up to $N=15$ and stretching ratio $d=1$. The general behavior is a 
rise turning over into an exponential decay. For small $\varepsilon$ the rise
is linear, in agreement with the results presented for the zero field 
diffusion coefficient. In fact we have calculated the diffusion coefficient as 
the derivative of the drift velocity with respect to $\varepsilon$ 
(see Sec. (\ref{subsec:diffusion})). The exponential decay is in accordance 
with the behavior predicted by Kolomeisky \cite{Kolothesis}. 
For large fields the probability distribution narrows down to a single $U$ 
shaped configuration with equally long  arms at both sides of the center of 
the chain. For odd chains Kolomeisky shows that, in the limit $\varepsilon 
\to \infty$ \cite{Kolothesis}
\begin{equation} \label{kolo}
v \sim \exp\left(\frac{2-N}{2} \varepsilon\right) 
\end{equation}
For the chains up to $N=7$ we could confirm this behavior (see inset of Fig.
\ref{FIG07}), but for the longer chains one would have to go to larger values 
of $\epsilon$ than can be handled with the Arnoldi method to see the asymptotic 
behavior. Generally the curves are a compromise between two tendencies: 
the expression for the drift grows with the field but the probability on a 
configuration shifts towards configurations with the least velocity.
With increasing $\epsilon$ initially the first effect dominates but gradually 
the second effect overrules the first.

\begin{figure}
\centerline{{\psfig{file=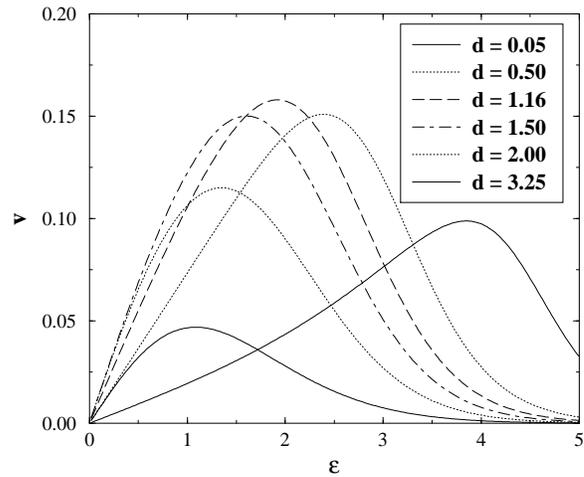,height=6.5cm}} }
\vskip 0.4truecm
\caption{The velocities as a function of an electric field for different $d$.
The number of reptons is fixed at $N=7$.} 
\label{FIG08}
\end{figure}

We have also investigated the behavior of the velocity as function of the 
stretching ratio $d$. In Fig. \ref{FIG08} we have plotted the various 
curves for chain length $N=7$ for $d$ ranging from $0.05$ to $3.25$. We see 
that the overall velocity becomes small for small $d$ which is simply a sign of 
slowing down due to the slowed down motion of the end reptons. For the higher 
values of $d$ we see again a decrease in the drift velocity, which is 
explained by stretching of the chain as we will see.

A closer inspection of the curves $v$ vs. $\varepsilon$ in the interval
$0 < \varepsilon \leq 1$ for the longer chains shows a definite deviation 
of the linear rise to larger values of the drift velocity. This interesting 
intermediate regime was also investigated by Barkema {\it et al.} \cite{Bark94}
by means of Monte Carlo simulations. They propose to describe the drift 
velocity in this region by the phenomenological crossover expression
\begin{equation} 
\label{a2}
v(\epsilon, N) \simeq \frac{\epsilon}{(2 d +1) N} [1 + 
A (\epsilon N)^2)]^\frac{1}{2}
\end{equation}
which seems to fit quite well the Monte Carlo data \cite{vanH00}.
Thus in the regime where $\epsilon$ is small but the combination $\epsilon N$ 
of order unity, the drift velocity becomes independent of $N$ (band collapse)
and quadratically dependent on $\epsilon$. 

\begin{figure}
\centerline{{\psfig{file=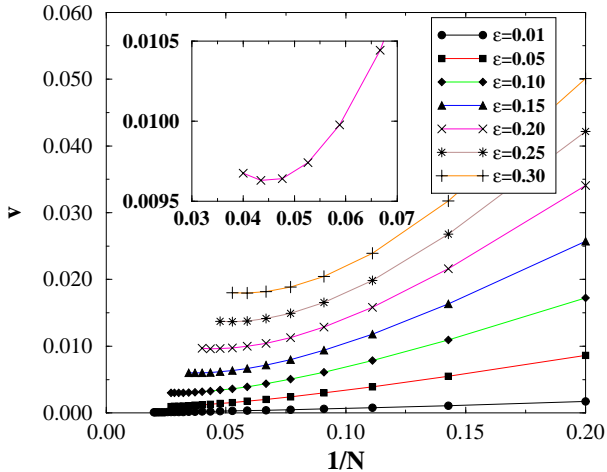,height=6.5cm}} }
\vskip 0.4truecm
\caption{Plot of the drift velocity $v$ as function of the inverse chain
length $1/N$, for various fixed values of the electric field $\varepsilon$
and $d=1$.
Inset: Blow up of $v$ vs. $1/N$ for $\varepsilon=0.2$; a minimum in the
velocity can be clearly seen.}
\label{FIG09}
\end{figure}

In Figure \ref{FIG09} we plotted the velocity as function of the inverse 
chain length $1/N$ for various values of the electric field. The deviations 
from the linear regime, where $v \sim \varepsilon/N$, are clearly observable 
as for sufficiently long chains and not too small fields the drift velocity 
approaches a non-vanishing value. 
A closer inspection on the curves reveals that the limiting constant velocity
in the asymptotic regime $N \to \infty$ is reached through a {\it minimum}
in the velocity at a given polymer length $N_{\rm min}$. In our calculations 
this occurs for any values of the fields in the range $\varepsilon \geq 0.10$. 
Presumably the minimum is shifted to lengths $N$ much longer than those which 
can be reached in the present calculation.
The existence of a minimum velocity at finite length can be clearly seen in the 
inset of Fig. \ref{FIG09} which plots $v$ vs. $1/N$ for $\varepsilon=0.2$, but 
it is present also in other curves. According to our error estimates typical 
uncertainties in the velocities are at the sixth decimal place, thus error bars 
are much smaller than all symbol sizes in the figure and inset.

The possible existence of a velocity minimum has given rise to quite some
discussions in the past and it is a phenomenon referred to as {\it band
inversion} (for a review see Ref.\cite{Viov00}). 
In the band inversion region long polymers migrate faster along the field 
direction than short ones, a physical situation that looks at first sight 
somewhat counterintuitive. Band inversion was predicted first in the context 
of the biased reptation model (BRM) \cite{Nool87}, which provides a simplified 
picture of the physics of polymer reptation in a field, in which fluctuations 
effects are neglected. More refined analytical calculations 
in models where fluctuation effects are taken into account still predict the
presence of a velocity minimum \cite{Seme95}, thus the original result of the 
BRM was confirmed.

It should be pointed out that the Eq. (\ref{a2}) does not yield any minimum 
in the velocity as $\partial v/\partial N < 0$ for all $N$, thus it cannot be
strictly correct. Although our results are limited to the onset of the minimum,
we suspect that the minimum is not a very pronounced one so that for
longer chains the curves $v$ vs. $1/N$ would appear rather flat.
A shallow minimum could have been thus missed in the Monte Carlo simulations
of Ref. \cite{Bark94}.

Indeed, the data presented in Ref. \cite{Bark94} yield, for instance, 
$v \approx 0.010$ for $\varepsilon = 0.2$ and $N=100$, thus the limiting 
velocity appears to be approached very slowly from below (see inset of 
Fig. \ref{FIG09}).
Other previous simulations by Duke \cite{Duke89} of the RD model yielded 
instead a quite clear velocity minimum, but it is difficult to judge the
quality of his data as no error bars are reported. Probably the inversion
effect is somewhat weaker than reported in \cite{Duke89}.
We investigated further the effect of $d$ and found that for higher $d$ the
minimum appears to be somewhat more pronounced, which may explain the results
of Ref. \cite{Duke89}, where $d=6$ was taken (the lattice coordination 
number for an fcc lattice).

Unfortunately, present DMRG computations are limited to the onset of the
inversion phenomenon.
The problem with the DMRG approach is that it builds up an optimal basis for
the stationary state using information of stationary states for shorter
chains. Around the band inversion point there is a kind of phase transition
from short unoriented polymers to long oriented ones (see for instance 
Ref. \cite{Viov00}). This change implies that the optimal basis for 
short chains may be no longer a good one when longer chains are considered.
We alleviated somewhat this problem using the DMRG algorithm at fixed 
$N \varepsilon$, which allows indeed to study slightly longer chains, 
yet not enough to go deep into the band collapse regime, but sufficient
to bring in evidence the velocity minimum.

Originally, it was thought that the drift velocity should be described by a
scaling function in terms of the combination $N \varepsilon^2$
\cite{Lump85,Slat86}. It is nowadays believed that the correct scaling form 
for the velocity should be given by an expression \cite{Duke92,Bark94}
\begin{equation}
v(\varepsilon,N) =\frac{\varepsilon}{N} \ g(N \varepsilon)
\label{sccc}
\end{equation}
with $g(x) \to g_0 > 0$ for $x \to 0$ and $g(x) \sim x$ for $x \to \infty$ in
order to match the known behavior of the velocity at large and small fields.
The condition to have band inversion is $\partial v/\partial N = 0$ for some 
$N$ at fixed fields, which is equivalent to the requirement $g(x) = x g'(x)$, 
for a non-zero value of $x = N \varepsilon$.
Notice that choosing the most general scaling form 
$v(\varepsilon,N) =\frac{\varepsilon}{N} \ g(N \varepsilon^\alpha)$
and from the requirement of $\partial v/\partial N = 0$ one still
obtains $g(x) = x g'(x)$, with $x = N \varepsilon^\alpha$. 

\begin{figure}
\centerline{{\psfig{file=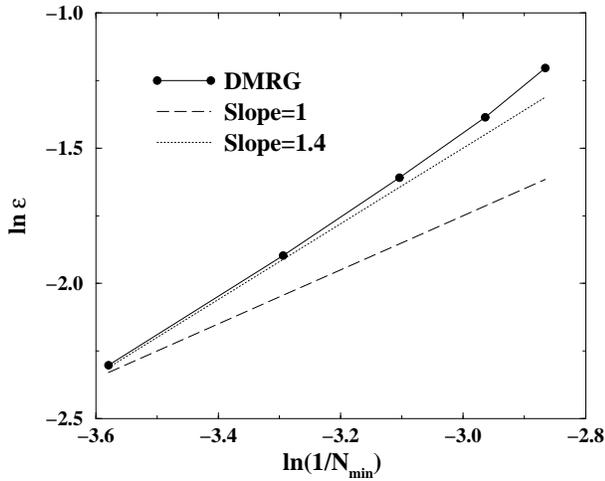,height=6.5cm}} }
\vskip 0.4truecm
\caption{Plot of $\ln \varepsilon$ vs. $\ln N_{\rm min}$ for $d=1$.}
\label{FIG10}
\end{figure}

The scaling behavior of the minimum as function of the field may be used to 
test the velocity formula.
We calculated the polymer length $N_{\rm min}$ at which the minimum of the
velocity occurs as function of the applied field $\varepsilon$. 
If Eq. (\ref{sccc}) is correct we expect $\varepsilon \sim 1/N_{\rm min}$.
Figure \ref{FIG10} shows a plot of $\ln \varepsilon$ vs.  $\ln N_{\rm min}$ 
for $d=1$. The data show some curvature due to corrections
to scaling and approach, for the longest chains analyzed, the asymptote 
$\varepsilon \sim N^{-\alpha}$, with $\alpha = 1.4$, as illustrated in the
figure. As $N_{\rm min}$ is increased one observes a systematic decrease of
the local slope of the data suggesting that the asymptotic exponent 
$\alpha < 1.4$.
In an attempt to reach the asymptotic regime we extrapolated the local slopes 
of the data in Fig. \ref{FIG10} in the limit $N_{\rm min} \to \infty$, which
yield an extrapolated value $\alpha \approx 1.1$, not far from the prediction 
of Eq.(\ref{sccc}), but not fully consistent with it. To prove Eq.(\ref{sccc})
more convincingly one would need to investigate longer chains.

\subsection{Correlations and profiles}
\label{sec:fieldb}

A more detailed insight in the shapes of the configurations is obtained by
plotting averages of the local variables $y_i$. The DMRG procedure leads
naturally to the determination of the probabilities for two consecutive 
segments
\begin{equation} \label{a3}
p_i (y,y') = \langle \, \delta_{ y , y_i}  \, \delta_{y' , y_{i+1}} \,
\rangle.
\end{equation}
The probabilities on a single segment follow from these values
\begin{equation} \label{a4}
p_i (y) = \sum_{y'} p_i (y,y') = \langle \, \delta_{ y , y_i}  \, \rangle =
\sum_{y'} p_{i-1} (y',y),
\end{equation}
which in turn are normalized
\begin{equation} \label{a5}
\sum_y p_i (y) = 1.
\end{equation}

The 9 possible values of $p_i (y,y')$ are restricted by these conditions.
One finds another set of relations between these quantities by summing the 
Master Equation over all but one segment value. Exclusion of an internal 
segment $y_i$ from the summation yields for $y_i=y=\pm 1$
\begin{eqnarray} \label{a6}
B^{y} p_{i-1} (0,y) - B^{-y} p_{i-1} (y,0) = \nonumber \\
B^{y} p_i (0,y) - B^{-y}  p_i (y,0)  = v(y).
\end{eqnarray}
Note that $v(y)$ is independent of the index $i$ of the segment under
consideration. The two relations (\ref{a6}) are an expression of the fact
the average velocity of the reptons in the field direction and the
curvilinear velocity are constant along the chain.  Taking the segment 
value $y_j=0$ one finds
\begin{equation} \label{a7}
v = v(1) - v(-1) = 2 v(1),
\end{equation}
with $v$ the drift velocity.
Excluding the end segments from the summation yields the 2 equations ($y =
\pm 1 $)
\begin{eqnarray} \label{a8}
v (y) = B^y p_1 (y) - d B^{-y} p_1 (0)  \nonumber \\
v (y) = dB^y p_N (0) - B^{-y} p_N (y)
\end{eqnarray}

One observes that the 3 probabilities on the end segments are fixed by the
normalization and the  2 equations (\ref{a8}). In general these relations 
show how delicate the development of the correlations is. In the fieldless 
case the probability of a configuration factorizes in a product over
probabilities of segments. So for $d=1$ the probability on any segment
becomes equal to 1/3. Then of course the velocities vanish. Considering the 
terms linear in the field (or in $B-B^{-1}$) one sees that the drift velocity 
of the first repton, given by
\begin{equation} \label{a9}
v = B p_1 (1) - B^{-1} p_1 (-1) + d (B - B^{-1}) p_1 (0),
\end{equation}
requires a delicate compensation in the linear deviations of the
probabilities $
p_1 (y)$
in order to give a value which vanishes as $1/N$ for long chains.

Rather than giving the values of $p_i (y)$ we plot the averages
\begin{equation} \label{a10}
\langle \, y_i \, \rangle = p_i (1) - p_i (-1)
\end{equation}
and \begin{equation} \label{a11}
\langle \, y_i \, y_{i+1} \,\rangle = p_i (1,1) + p_i (-1,-1) - p_i (1,-1) -
p_i
 (1-,1).
\end{equation}

\begin{figure}
\centerline{
{\psfig{file=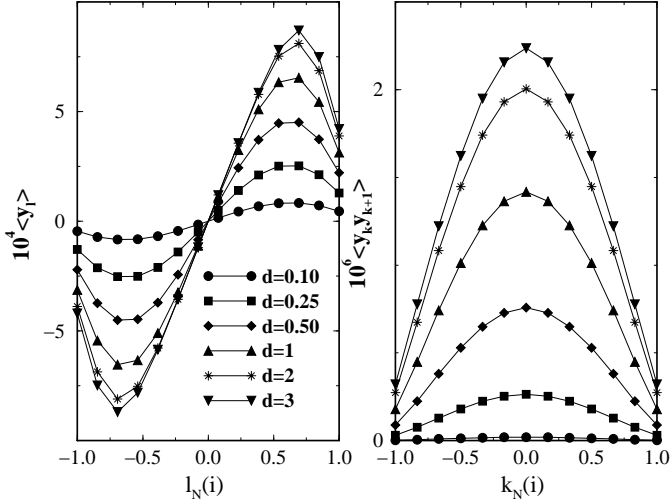,height=7.0cm}}
}
\vskip 0.4truecm
\caption{The profiles for $\varepsilon=0.001$ plotted as function of the reduced
distances $l_N(i) = (2i-N)/(N-2)$ and $k_N(i) = (2i-N+1)/(N-3)$ 
(with $i = 1,2, \ldots N-1$).
The number of reptons is fixed at $N=15$.
}
\label{FIG11}
\end{figure}

\begin{figure}
\centerline{ {\psfig{file=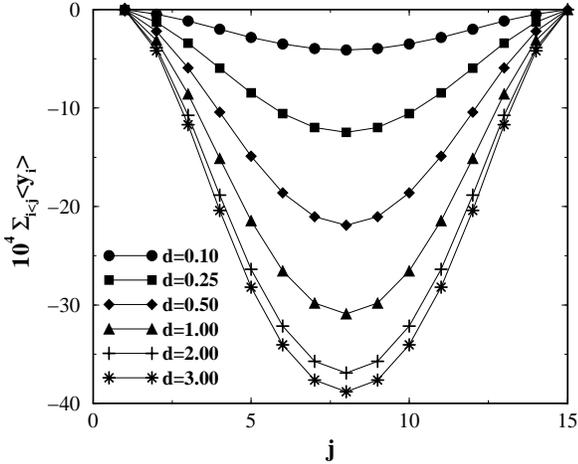,height=7.0cm}} }
\vskip 0.4truecm
\caption{The shapes of the chain for $\varepsilon=0.001$ and various 
values of $d$. The number of reptons is fixed at $N=15$.
}
\label{FIG12}
\end{figure}

A typical plot for the small $\epsilon$ regime is given in Fig. \ref{FIG11}. 
The global symmetry due to the interchange of head and tail makes the average 
(\ref{a10}) anti--symmetric with respect to the middle and the average 
(\ref{a11}) symmetric. One observes that the features increase with the 
mobility $d$ of the end reptons. The averages give information about the 
average shape of the chain. 
In Fig. \ref{FIG12} we translate the averages (\ref{a10}) 
into average spatial configurations by integrating (summing) the segment 
values to positions with respect to the middle repton. Clearly the development 
of the $U$ shape is visible with increasing $d$. The effect is larger at the 
ends than in the middle.

\begin{figure}
\centerline{
{\psfig{file=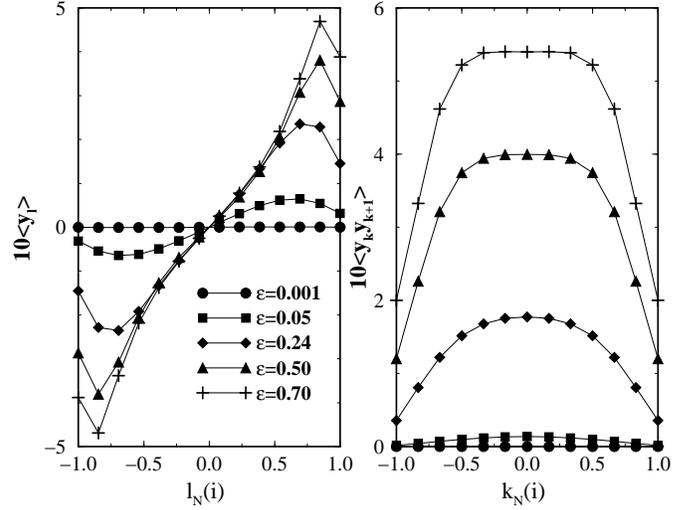,height=7.0cm}}
}
\vskip 0.4truecm
\caption{As in Fig. \ref{FIG11} for $d=1$ and varying $\varepsilon$.
The number of reptons is fixed at $N=15$.}
\label{FIG13}
\end{figure}

As an example of the behavior in the intermediate regime (of the velocity
profiles Fig. 3) we have plotted in Fig. \ref{FIG13} the situation for $N=15$, 
$d=1$ and various values of $\epsilon$. For the larger values of $\epsilon$ 
the average $\langle y_i \rangle$ does not change very much but the plateau 
of the correlations $\langle y_i y_{i+1} \rangle$ in the middle keeps rising. 
We expect that for longer chains a large region in the middle develops where 
the shape varies weakly with $\epsilon$ and where the correlations between 
consecutive segments increase. Thus the chain obtains longer stretches which 
are oriented in the field, either up or down, but which largely compensate, 
such that the overall shape in the middle remains more or less the same. 
This is in agreement with the speculations of Barkema {\it et al.}
\cite{Bark94} on the chain as a stretched sequence of more or less isotropic 
"blobs".

\begin{figure}
\centerline{
{\psfig{file=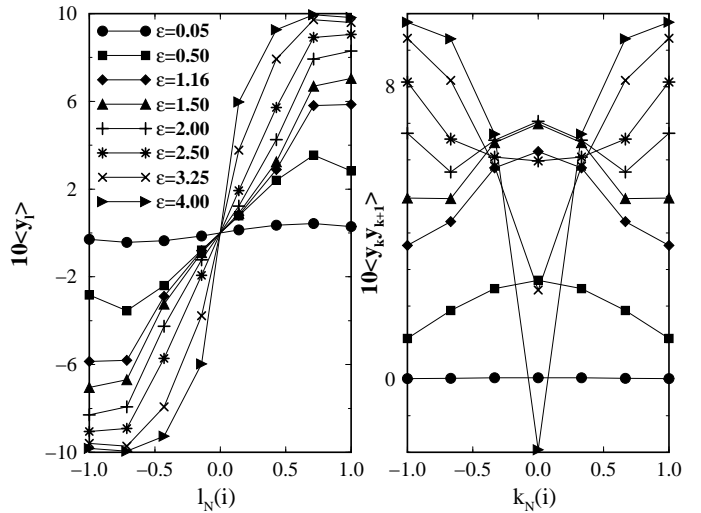,height=7.0cm}}
}
\vskip 0.4truecm
\caption{The profiles for $d=1$. 
The number of reptons is fixed at $N=9$.
}
\label{FIG14}
\end{figure}

Finally we show in Fig. \ref{FIG14} the situation for strong $\epsilon$ on a 
chain of $N=9$ reptons. For the strongest values of $\epsilon$ it is almost
exclusively in the $U$ shaped configuration. Note also that the correlations
approach 1 except for the middle pair of segments which are on different
branches of the $U$. So ultimately the value of the correlation in the
middle will approach -1.

Thus we see that the reptating chain develops a rich and delicate pattern
of shapes and correlations which are not so easy to catch in simple 
describing formulae. The difficulty is that there is a different
dependence in scale on the parameters $\epsilon $ and $N$ in the
middle of the chain and at the ends of the chain. A rough estimate
indicates the existence of a zone of length $\sqrt{N}$ at the ends of
the chain with the typical bending over of the average $\langle y_i \rangle$.
In the middle remains a zone of length order $N$, with 
fairly constant correlations. This splitting up in ``bulk'' and ``surface''
behavior which keep each other in balance prevents a systematic expansion
in the small parameter $\epsilon$. 

\section{Discussion}
\label{sec:discussion}

Using the DMRG technique we have determined the properties of the RD model
for moderately long chains. At zero and for very small driving fields we could 
reach chains of the order of $N \approx 100-150$ reptons; for finite fields 
the lengths are restricted to some 30 reptons. This regime is far outside the 
domain where exact diagonalization of the reptation matrix is possible. 
The DMRG 
results have the advantage, over corresponding Monte Carlo results, of being virtually 
exact as long as the iteration method converges. Since the DMRG procedure gives 
simultaneously all the lengths $N$ smaller than the maximum, we can accurately 
determine the finite size effects on the asymptotic large $N$ behavior.

We find that the renewal time $\tau$ and the diffusion coefficient $D$ are 
strongly affected by finite size corrections in the regime of these moderately 
large $N$. 
Here we have shown that the large finite size corrections, characteristic 
for the reptation process, manifest themselves as effective exponents for 
the asymptotic behavior of the renewal time and the diffusion coefficient.
We found that DMRG results reveal that, while the leading correction terms, 
as given by Doi's theory \cite{Doi81} fit rather well the data for large $N$, 
they are not sufficient to cause a crossover behavior and higher order 
corrections need to be included. 
These finite size effects offer an explanation for the discrepancies between 
measurement and standard theory. This point can be further tested 
experimentally by playing with the structure of the end reptons (e.g. 
branching) and the coordination number of the embedding lattice. We have 
lumped these aspects in a ``dimensionality" parameter $d$, which indeed 
has surprising effects on the finite size behavior.
In particular we expect that chains with short branching ends will be 
mapped onto small $d$ regimes, where $\tau$ and $D$, according to our 
results, will scale with effective exponents deviating from the values
measured so far. Experimental tests of this prediction will possibly 
provide new insight for the understanding of the dynamics of entangled 
polymer melts and concentrated solutions.

We have also established the onset of the so--called band inversion. For 
fixed driving field and increasing $N$ we observe that the drift velocity 
goes through a minimum.
This is another intriguing effect contained in the RD model. The band 
inversion has been ascribed to the fact that long polymers in a driving 
field get oriented and that due to this fixed orientation the drift velocity 
rather increases with $N$ than decreases as in the non--oriented regime. 
To see the ultimate asymptotic velocity longer chains than presently possible 
should investigated.

The DMRG calculations also yield a host of detailed information about the 
structure of the reptating polymer as for instance the local correlation 
functions. We have plotted the average values $\langle y_i \rangle$ and 
$\langle y_i y_{i+1} \rangle$.
We find that the reptating chain develops a rich and delicate pattern
of shapes and correlations which are not so easy to catch in simple 
describing formulae. The difficulty is that there is a different
dependence in scale on the parameters $\epsilon $ and $N$ in the
middle of the chain and at the ends of the chain. A rough estimate
indicates the existence of a zone of length $\sqrt{N}$ at the ends of
the chain with the typical bending over of the average $\langle y_i \rangle$.
In the middle remains a zone of length order $N$, with 
fairly constant correlations. This splitting up in ``bulk'' and ``surface''
behavior which keep each other in balance prevents a systematic expansion
in the small parameter $\epsilon$. 

We are grateful to G. T. Barkema for helpful discussions and to T. Lodge for
drawing our attention to the experimental results of Refs. \cite{Lodg99}
and \cite{Thao00}.

\end{multicols}
\end{document}